\documentclass[sigconf,nonacm]{acmart}

\AtBeginDocument{%
  }

\newcommand{\workshopname}{FDG 2024 Game Demos: \url{http://fdg2024.org/documents/FDG_2024_PAPERS.pdf}}
\copyrightyear{2024}
\setcopyright{rightsretained}
\newcommand\extrafootertext[1]{
    \bgroup
    \renewcommand\thefootnote{\fnsymbol{footnote}}%
    \renewcommand\thempfootnote{\fnsymbol{mpfootnote}}%
    \footnotetext[0]{#1}%
    \egroup
}

\AtBeginDocument{ 
    \fancypagestyle{firstpagestyle}{
        \fancyhf{}
        \fancyfoot[L]{\sffamily\footnotesize \workshopname}%
        \fancyfoot[R]{\sffamily\footnotesize \thepage}
    }
    \fancyhf{}
    \fancyhead[L]{\sffamily\footnotesize\shorttitle}
    \fancyhead[R]{\sffamily\footnotesize\shortauthors}
    \fancyfoot[L]{\sffamily\footnotesize\workshopname}%
    \fancyfoot[R]{\sffamily\footnotesize\thepage}
}

\acmDOI{}
\acmISBN{}

\begin{document}

\title{Info Overload: A Cooperative Evacuation Game}

\author{MJ Johns}
\email{mljohns@ucsc.edu}
\orcid{0009-0002-4016-2517}
\affiliation{%
  \institution{University of California Santa Cruz}
  \streetaddress{1156 High St}
  \city{Santa Cruz}
  \state{California}
  \country{USA}
  \postcode{95064}
}

\author{Rita Tesfay}
\email{rtesfay@ucsc.edu}
\affiliation{%
  \institution{University of California Santa Cruz}
  \streetaddress{1156 High St}
  \city{Santa Cruz}
  \state{California}
  \country{USA}
  \postcode{95064}
}
\author{Mário Escarce Junior}
\email{m.escarce@lancaster.ac.uk}
\affiliation{%
  \institution{Lancaster University}
  \streetaddress{LA1 4YW}
  \city{Lancaster}
  \country{UK}
}
\author{Emmanuel Ezenwa Jr.}
\email{eezenwa@ucsc.edu}
\affiliation{%
  \institution{University of California Santa Cruz}
  \streetaddress{1156 High St}
  \city{Santa Cruz}
  \state{California}
  \country{USA}
  \postcode{95064}
}
\author{Thomas Maiorana}
\email{tmaiorana@ucdavis.edu}
\affiliation{%
  \institution{University of California Davis}
  \streetaddress{1 Shields Ave}
  \city{Davis}
  \state{California}
  \country{USA}
  \postcode{95616}
}

\author{Magy Seif El-Nasr}
\email{mseifeln@ucsc.edu}
\affiliation{%
  \institution{University of California Santa Cruz}
  \streetaddress{1156 High St}
  \city{Santa Cruz}
  \state{California}
  \country{USA}
  \postcode{95064}
}
\author{Edward Melcer}
\email{eddie.melcer@ucsc.edu}
\affiliation{%
  \institution{University of California Santa Cruz}
  \streetaddress{1156 High St}
  \city{Santa Cruz}
  \state{California}
  \country{USA}
  \postcode{95064}
}

\author{Katherine Isbister}
\email{katherine.isbister@ucsc.edu}
\affiliation{%
  \institution{University of California Santa Cruz}
  \streetaddress{1156 High St}
  \city{Santa Cruz}
  \state{California}
  \country{USA}
  \postcode{95064}
}

\renewcommand{\shortauthors}{Johns et al.}

\begin{abstract}
  Info Overload is a minigame within a larger collection aimed at increasing awareness and preparation for an evacuation in the event of a wildfire. The game relies on experiential two-player cooperative gameplay and is played on a mobile device (a phone or tablet). 
\end{abstract}



\maketitle

\section{Introduction}
In the event of a disaster, such as wildfire, many neighborhoods and communities in the path of destruction will be ordered to evacuate to save as many lives as possible. While community leaders and experts provide as much accurate direction as they can, the individuals evacuating are likely to face unforeseen obstacles and the need to make quick decisions with limited or conflicting information. 

We have designed the Info Overload minigame – part of a suite of minigames addressing wildfire preparation – to help people understand the evacuation process and the challenging situations they may need to think about. 

The game is cooperative for two players sharing one device (a phone or tablet) to simulate the experience of evacuating with another person in the same vehicle. In an effort to reduce congestion along evacuation routes, multicar households are strongly encouraged to evacuate in one vehicle and leave the other behind. The game emphasizes communication between the two players.

\section{Related Work}
Some existing games address wildfire preparedness, including This is Fine \cite{larakaa_this_2020} which is a game about putting out fires within your home, Disaster Master \cite{ready_disaster_2024} which is an interactive comic about staying alert for signs of wildfire, and Stop Disasters \cite{playerthree_stop_2018} which is a top-down simulation management game about creating safer communities that can withstand various disasters (including wildfire) from a city planning perspective. 

What sets ours apart is the perspective of a first-person view within the car during an active evacuation, the emphasis on cooperation and communication between two players working together, and the real-time decision making based on changing information. Outside of climate resilience, two entertainment-focused games exemplify the real-time cooperative gameplay we explore in Info Overload. Space Team \cite{henry_smith_spaceteam_2012} is a local-multiplayer mobile game where different players have different information which they must relay to each other to accomplish collaborative tasks to control a spaceship. Keep Talking and Nobody Explodes \cite{steel_crate_games_keep_2018} similarly involves collaborative tasks, except in this case one player wears a VR headset while the other players have a manual of instructions that must be relayed to help the VR player diffuse a bomb. 

\section{Design}
\begin{figure}[ht]
    \centering
    \includegraphics[width=0.9\linewidth]{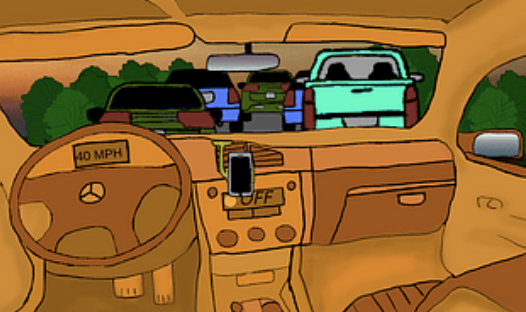}
    \caption{Screenshot of Gameplay}
    \label{fig:gameplay}
\end{figure}

Info Overload is designed to simulate the experience of two people evacuating in a car during a wildfire, as can be seen in figure \ref{fig:gameplay}. The players each have their own responsibilities and must work together to safely evacuate. 

Player one is driving and has control of the steering wheel and brakes. When they see brake lights ahead of them or emergency flashing lights in the rearview mirror they should hold the brake button down to bring the car to a stop. Decisions about when to turn require directions from player two. 

Player two is in charge of triaging information and navigating to safety. They have a phone with GPS as well as alert messages. They may also choose to turn on the radio for additional information. The text alerts will give them updated information about their route, such as unsafe roads or the need to find temporary shelter if the fire is moving quickly. The map view has a marker for their designated exit point, and they use the information from alerts to decide the best places to turn, which they relay to player one. 

The car is constantly moving forward (except when player one uses the brakes) and can only turn at intersections. If the players reach a deadend or fail to heed a warning about finding temporary shelter, they will be prompted to start over as the ‘lose’ condition. If they safely reach the designated exit point, they win the minigame. 

Our objective with the design is to avoid overly harsh penalties or making the players feel overwhelmed or upset, but to still convey the importance of communicating with others, using caution during an evacuation, and paying attention for updated information about their route. 
\section{Playable Demo}

\begin{figure}[h]
    \centering
    \includegraphics[width=0.35\linewidth]{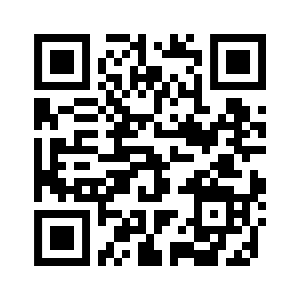}
    \caption{QR Code to play the game demo}
    \label{fig:qr-code}
\end{figure}

Play the game here: \url{https://ucsc-wildfire-games.itch.io/info-overload} or by scanning the QR code in Fig. \ref{fig:qr-code}. 

\section{Early Feedback}
The game is still in early development, but we have had some opportunities to discuss the design with community members and local experts at Participatory Design workshops. 

\begin{figure}[ht]
    \centering
    \includegraphics[width=0.9\linewidth]{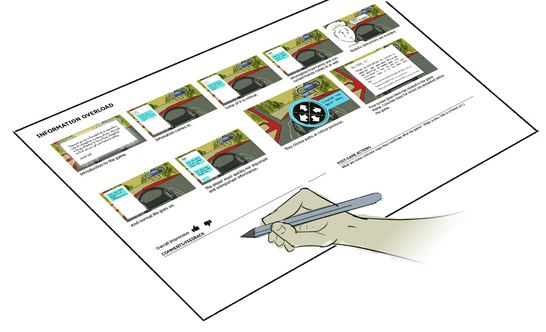}
    \caption{Initial storyboard shown at the first workshop}
    \label{fig:form}
\end{figure}

\begin{figure}[ht]
    \centering
    \includegraphics[width=1\linewidth]{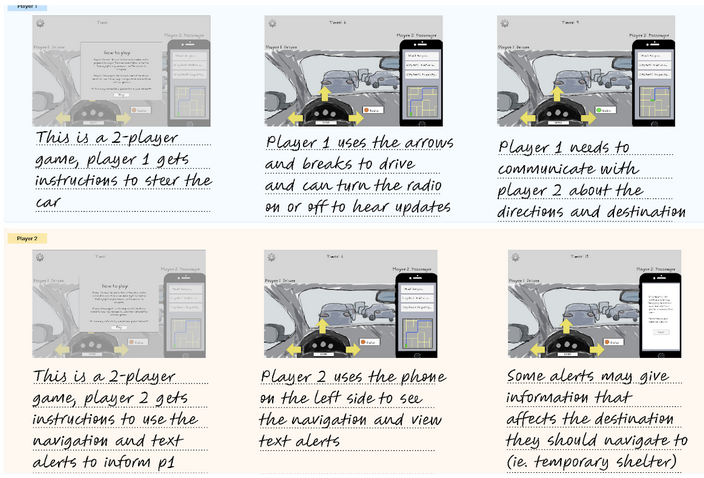}
    \caption{Revised storyboard used at second workshop}
    \label{fig:storyboard}
\end{figure}

Participants suggested we incorporate observable contextual information, such as stop lights being out or seeing smoke or fire in the distance. In general they liked that we had a two-player experience so players could practice working together, but they had some concerns about players being frustrated with each other, particularly if they have different levels of skill. 

Some workshop participants with fire training also pointed out that a lot of this scenario in real life would be outside of the players control, and cautioned us to avoid giving a false sense to the player of how much they would realistically be able to do about it. 

We also learned that while some communities rely on radio for real-time updates, others do not, so we may need to consider regional variants for our design. 

\section{Conclusion and Future Work}

Info Overload is a minigame to improve preparedness for wildfire evacuation by simulating the experience of evacuating in a vehicle, allowing players to practice communication and collaboration as well as triaging information during a disaster. As we continue to develop the game we will seek community feedback and insights from experts to ensure the information we present is accurate and beneficial.

\begin{acks}
This material is based upon work supported by the National Science Foundation under Grant No. 2230636. We also would like to thank our collaborators Dr. Kenichi Soga and Dr. Louise Comfort and their respective teams, and the community members who offered valuable insights at our community workshops. 
\end{acks}
\bibliographystyle{ACM-Reference-Format}
\bibliography{info-overload-authordraft}


\begin{thebibliography}{5}


\ifx \showCODEN    \undefined \def \showCODEN     #1{\unskip}     \fi
\ifx \showDOI      \undefined \def \showDOI       #1{#1}\fi
\ifx \showISBNx    \undefined \def \showISBNx     #1{\unskip}     \fi
\ifx \showISBNxiii \undefined \def \showISBNxiii  #1{\unskip}     \fi
\ifx \showISSN     \undefined \def \showISSN      #1{\unskip}     \fi
\ifx \showLCCN     \undefined \def \showLCCN      #1{\unskip}     \fi
\ifx \shownote     \undefined \def \shownote      #1{#1}          \fi
\ifx \showarticletitle \undefined \def \showarticletitle #1{#1}   \fi
\ifx \showURL      \undefined \def \showURL       {\relax}        \fi
\providecommand\bibfield[2]{#2}
\providecommand\bibinfo[2]{#2}
\providecommand\natexlab[1]{#1}
\providecommand\showeprint[2][]{arXiv:#2}

\bibitem[{Henry Smith}(2012)]%
        {henry_smith_spaceteam_2012}
\bibfield{author}{\bibinfo{person}{{Henry Smith}}.} \bibinfo{year}{2012}\natexlab{}.
\newblock \bibinfo{title}{Spaceteam ({Game})}.
\newblock
\newblock
\urldef\tempurl%
\url{http://spaceteam.ca/}
\showURL{%
\tempurl}


\bibitem[{Larakaa}(2020)]%
        {larakaa_this_2020}
\bibfield{author}{\bibinfo{person}{{Larakaa}}.} \bibinfo{year}{2020}\natexlab{}.
\newblock \bibinfo{title}{This is {Fine} ({Game})}.
\newblock
\newblock
\urldef\tempurl%
\url{https://larakaa.itch.io/this-is-fine}
\showURL{%
\tempurl}


\bibitem[{playerthree} and {UNDRR}(2018)]%
        {playerthree_stop_2018}
\bibfield{author}{\bibinfo{person}{{playerthree}} {and} \bibinfo{person}{{UNDRR}}.} \bibinfo{year}{2018}\natexlab{}.
\newblock \bibinfo{title}{Stop {Disasters} (game)}.
\newblock
\newblock
\urldef\tempurl%
\url{https://www.stopdisastersgame.org/}
\showURL{%
\tempurl}


\bibitem[{Ready}(2024)]%
        {ready_disaster_2024}
\bibfield{author}{\bibinfo{person}{{Ready}}.} \bibinfo{year}{2024}\natexlab{}.
\newblock \bibinfo{title}{Disaster {Master} (game)}.
\newblock
\newblock
\urldef\tempurl%
\url{https://www.ready.gov/kids/games/data/dm-english/wildfire.html}
\showURL{%
\tempurl}


\bibitem[{Steel Crate Games}(2018)]%
        {steel_crate_games_keep_2018}
\bibfield{author}{\bibinfo{person}{{Steel Crate Games}}.} \bibinfo{year}{2018}\natexlab{}.
\newblock \bibinfo{title}{Keep {Talking} and {Nobody} {Explodes} ({Game})}.
\newblock
\newblock
\urldef\tempurl%
\url{https://keeptalkinggame.com/}
\showURL{%
\tempurl}


\end{thebibliography}

\end{document}